\documentclass[pra,twocolumn,showpacs,amsmath,amssymb]{revtex4}

\usepackage{graphicx}
\usepackage{bm}

\newcommand{\rb}{$^{87}$Rb}

\begin{document}

\title{Spinor Dynamics-Driven Formation of a Dual-Beam Atom Laser }

\author{N. Lundblad}\email{lundblad@caltech.edu}
\altaffiliation[Also at ]{Physics Department, California Institute of Technology, Pasadena CA 91125 USA}

\author{R. J. Thompson}
\author{D. C. Aveline}
\author{L. Maleki}
\affiliation{Jet Propulsion Laboratory, California Institute of Technology, 4800 Oak Grove Drive, Pasadena, California 91109-8099, USA}
\date{\today}

\begin{abstract}
We demonstrate a novel dual-beam atom laser formed by outcoupling oppositely polarized components of an $F=1$ spinor Bose-Einstein condensate whose Zeeman sublevel populations have been coherently evolved through spin dynamics.    The condensate is formed through all-optical means using a single-beam running-wave dipole trap.    We create a condensate in the field-insensitive $m_F=0$ state, and drive coherent spin-mixing evolution through adiabatic compression of the initially weak trap.  Such dual beams, number-correlated through the angular momentum-conserving reaction 
$2m_0\leftrightharpoons m_{+1}+m_{-1}$,  have been proposed as tools to explore entanglement and squeezing in Bose-Einstein condensates, and have potential use in precision phase measurements.

\end{abstract}
\maketitle

Since the creation of the first spinor Bose-Einstein condensate (BEC) in 1998\cite{kettspinor}, a considerable body of work has emerged focusing on the properties and dynamics of such condensates, in which the spin degree of freedom has been liberated and the order parameter is vectorial.    Metastable states, spin domains, and coreless vortices were observed in sodium\cite{kettspinor2,kettspinor3,kettspinor4}, the  dynamics of spin mixing in both hyperfine ground states of rubidium have been extensively studied\cite{chapmanspinor,hamburg_f2,hamburg_f1,hamburg_revivals,japan_f2}, and the coherence of the mixing process has been concretely established\cite{chapmannature}.  Spin mixing has been used to drive the creation of multicomponent condensates at constant temperature\cite{hamburg_constantT}, the spatial magnetization profile of an $F=1$ spinor BEC has been observed\cite{stamperkurn}, and the clock transition has been explored\cite{kettspinor5}.   Considerable theoretical work has been devoted to the understanding of the spin-spin interaction Hamiltonian and the nature of the spinor condensate ground state\cite{ho,njp_spinor,lawpubigelow1998,japantheory,chapmantheory,bigelow1999}.      In addition, the spinor condensate has stimulated theoretical proposals regarding schemes to create squeezed and entangled beams of atoms via coherent spin mixing\cite{duan2000,pumeystre2000}.    Notions of entanglement and squeezing in dual-beam atom lasers are inherently related to the number conservation between the two evolved components, resulting in an Einstein-Podolsky-Rosen relationship between the measured number of each population.    In particular, these proposals suggested a scheme using spin-dependent light shifts to create oppositely propagating beams that were each superpositions of $m_F=\pm1$, and explored the correlations between them\cite{duan2000,pumeystre2000}.

In this article we present observations with an experimental scheme aimed at exploring these ideas: the generation of dual atom laser beams with an inherent number correlation between them due to their spin-mixing origin.    The novelty of the scheme lies not in the output coupler (simple magnetic field gradients that tilt the optical potential for the polarizable states) but rather in the  correlated nature of  the outcoupled populations and the utilization of spinor dynamics  to `pump' population into the outcoupled states.     In addition, the presence of a true reservoir and the ability to control both the strength of the output coupler and the rate of `pumping' make this a particularly intriguing atom-laser scheme.

  In an all-optical experiment, the path to condensation is largely unconcerned with spin, yet once degeneracy (as well as the high densities associated with it) sets in, the spin-spin interactional energy scale becomes relevant.    As such, the interparticle interaction can be described by $U({\mathbf r})=\delta(\mathbf{r})(c_0+c_2 \mathbf{F}_1\cdot \mathbf{F}_2)$, where $c_0$ is the traditional scattering length defined as $c_0 = (4\pi\hbar^2/m)(a_0+2a_2)/3$, and $c_2$ is the spin-spin energy scale, which is given by $c_2 = (4\pi\hbar^2/m)(a_2-a_0)/3$\cite{ho}.    Here $a_2$ and $a_0$ are the well-known \rb\  scattering lengths in the total spin channels $f=2$ and $f=0$, respectively\cite{ferro1,ferro2}.

It has been established that the spin-spin interaction of \rb\  is ferromagnetic\cite{chapmanspinor,hamburg_f2}, i.e. $c_2<0$, yet since the energy scale is so small ($c_2=-3.6\times10^{-14}$ Hz cm$^3$) a true ground state of this new Hamiltonian is not observable except in $\mu$G-level magnetically shielded environments\cite{chapmanspinor}.   Nevertheless, a consequence of this interaction at finite field is that spin mixing will occur; in particular, a pure condensate in the field-insensitive $m_F=0$ sublevel will coherently evolve into a superposition condensate through the collision $2m_0\leftrightharpoons m_{+1}+m_{-1}$, which for the condensate atoms is the only allowed spin dynamics\cite{ho,chapmanspinor}.    This spin mixing should be first observed after a timescale of order $\tau_0=1/(c_2 n)$ at zero field, where $n$ is the mean condensate density; however, the quadratic Zeeman shift is opposed to this evolution, and at a magnetic field where any evolution away from $m_F=0$ is energetically unfavorable we expect no mixing to occur.    In addition, despite low fields, if the condensate number is low enough such that the Thomas-Fermi densities force $\tau(B^2)= 1/(c_2 n-\nu_{B^2})$ to be significant compared to the condensate lifetime, mixing will be inaccessible (this restriction is overcome through adiabatic compression, as detailed below).         Worth particular mention among the many other investigations of spin dynamics are the observations  of number correlation in the evolved components, performed by examining fluctuations in the $\pm 1$ populations of mixed condensates\cite{chapmanspinor}.     Also notable is the observation of number-correlated atom laser beams generated as a result of four-wave mixing in a sodium BEC\cite{kett4wm}.

Our apparatus, similar to that of other groups\cite{tubingen,chapman}, utilizes a single-beam running-wave dipole trap produced by a focused CO$_2$ laser, which provides at full power a trap depth of approximately 1.6 mK via the DC polarizability of \rb.    We load the dipole trap from an ultra-high vacuum magneto-optical trap (MOT) which is itself loaded by a cold atomic beam provided by an upstream 2-dimensional MOT\cite{2dmot_amsterdam}.  The 2D-MOT exists in a rubidium vapor cell which is differentially pumped from the adjoining science chamber.    All 780nm trapping light is provided by a laser system based on a frequency-doubled 1560nm fiber amplifier, described elsewhere\cite{doubling}.    The loading of the dipole trap proceeds according to established technique\cite{chapman}; we obtain initial populations in the trap of $\sim 2\times10^6$ \rb\ atoms at $\sim 120\  \mu K$.   The initial trap frequencies (measured via parametric resonance) are approximately 3.2 kHz transversely and 220 Hz longitudinally.

Evaporative cooling proceeds via a programmed rampdown of CO$_2$ laser intensity.    We observed the onset of BEC at critical temperatures near 100 nK with around $10^5$ atoms, and typically obtain condensates of $3\times 10^4$ atoms with little or no discernible thermal component.   Typical trap frequencies for holding a formed condensate were 175 Hz (transverse) and 12 Hz (longitudinal).

\begin{figure}
\includegraphics[width=3.375in]{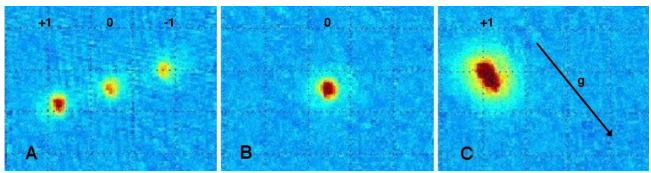}
\caption{\label{becpalette}  Spinor BEC creation options: a) the default triplet, with repeatable population distribution likely set by initial MOT alignment; b) $m_F=0$ trap, created by  selective application of magnetic field gradient along the weak axis of the trap; c) enhanced $m_F=+1$, created via application of a supportive gradient throughout evaporation.  Gravity is directed toward the lower right, and the trapping laser is directed toward the upper right.    All images are of partially condensed samples at a ballistic expansion time of 17.5 ms.   The long axis of the dipole trap  is directed toward the upper right.   Images are 1.3 mm $\times$ 1 mm. }
\end{figure}

The spin projections of the condensate are obtained through Stern-Gerlach separation during ballistic expansion.  The conveniently located MOT coils are used for this purpose.     Absorption imaging is performed on the expanded condensates  (or, in the case of the atom lasers, immediately after dipole trap turnoff) using the $F=2\rightarrow F'=3$ transition after optically pumping from the lower ground state.

Fig.~\ref{becpalette} summarizes the various types of dipole-trap BECs available to us in the current configuration.   Typical evaporation yields a condensate with all three components visible; the ratio of these three populations appears to be a constant of the experiment.   It has been speculated that this initial population is set by the particular location of the dipole trap within the MOT reservoir during trap loading\cite{lett}.

Application of a magnetic field gradient along the weak axis of the trap during the first few seconds of evaporation preferentially biases out the $m_F=\pm 1$ components, resulting in a BEC solely occupying the field-insensitive $m_F=0$ projection.   It should be noted that this process results in nearly the same number of condensed atoms as the gradient-free process, implying a sympathetic cooling process whereby the polarized components remove more than their average share of thermal energy.

Finally, application of a small magnetic field gradient of order a few G/cm in the vertical direction provides a bias for one or the other polarized components.   If this supportive gradient is only on for the first few seconds of evaporation, we obtain polarized condensates of number similar to the other options.   However, if this gradient is maintained through condensate formation and through ballistic expansion, we observe significant enhancement in condensate number of order 100\%; we attribute this to the fact that the trap depth is strongly perturbed by gravity near criticality, and even a gradient small with respect to gravity allows for much more efficient near-critical evaporation.

\begin{figure}
\includegraphics[width=3.375in]{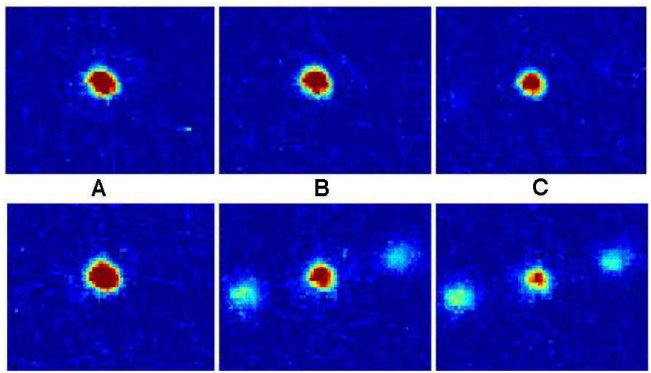}
\caption{\label{nomixing} The effect of adiabatic compression on the spin mixing process; the top row shows condensates held for equivalent durations without compression.  a) 100 ms of hold time at 700 mW b) 400 ms of hold time  c) 1.2 s of hold time.    The slightly fewer overall number in (c) is due to condensate lifetime.   The ballistic expansion time for all images is 17.5 ms.  Images are 1 mm $\times$ .8 mm.  }  
\end{figure}

For the observations reported here, we begin with a nominally pure $m_F=0$ condensate held in a trap whose unperturbed depth is 5 $\mu$K and is approximately a factor of 10 weaker due to gravitational tilt.    Typical condensate densities at this point are in the range $4-6\times 10^{13}$ cm$^{-3}$, and even given up to two seconds of observation spin mixing does not occur.    We fix background field levels at 60 mG as determined by RF spectroscopy.    To coherently mix the condensate we adiabatically compress the trapping field by raising the laser power from 100  mW to 700 mW over 100 ms, holding the compressed condensate for a variable time, adiabatically expanding, and finally ballistically expanding while applying the Stern-Gerlach field, as summarized by Fig.~\ref{nomixing}.    We observe that the fraction of atoms evolved into the polarized projections increases with high-density hold time and eventually reaches a static level of 50\%, as expected given previous experiments and theoretical prediction\cite{chapmanspinor,njp_spinor}.    Oscillations of spin populations are obscured by imaging noise  and considerable shot-to-shot number variance.    We also observe that the time taken to reach this steady state varies linearly with density, with a possible offset given by the density at which the quadratic Zeeman effect dominates the dynamics.     Densities are calculated in the Thomas-Fermi picture wherein the peak condensate density varies as $n_p\propto N^{2/5}U_0^{3/5}$, where $U_0$ is calculated from the trap frequencies alone and is only slightly perturbed by the strong gravitational tilt which so strongly affects evaporation.     We observe strong losses at high ($>$ 2 W) compression assumed to be caused by three-body recombination; the loss rates are consistent with projected values of the Thomas-Fermi densities and the measured value of the three-body rate constant $K_{3c}$\cite{burt}.

\begin{figure}
\includegraphics[width=3.375in]{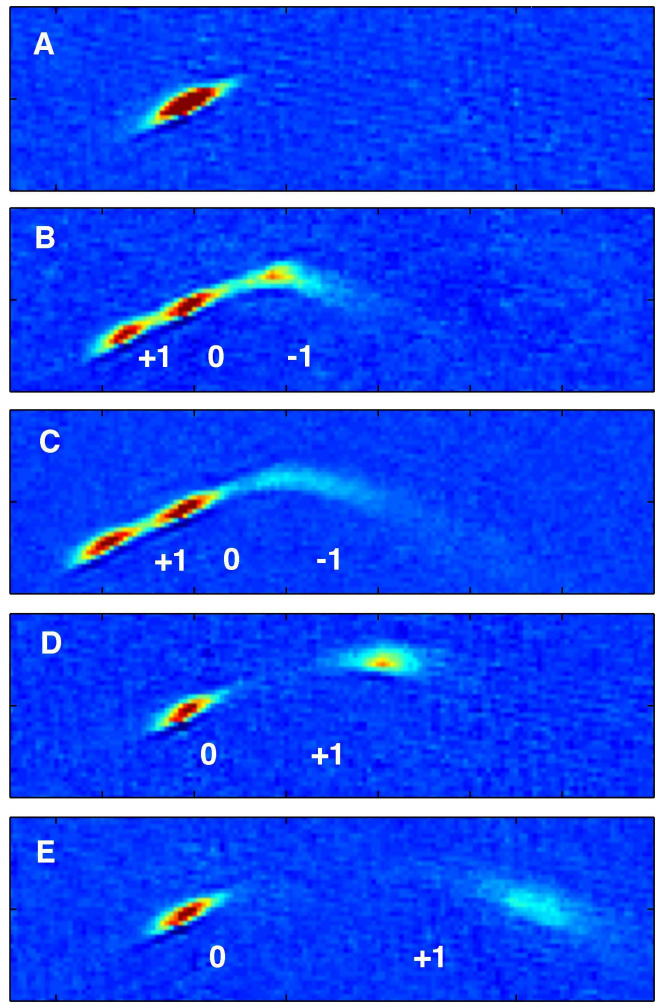}
\caption{\label{twinbeam}  A typical outcoupling run of the spinor dynamics-driven dual beam atom laser.  a) 0 ms: the full condensate, \textit{in situ}.   b)+ 20 ms: immediately after outcoupling.   The $m_F=-1$ component immediately passes beyond the reach of the dipole trap and experiences ballistic flight and mean-field expansion.  The $m_F=+1$ component remains confined in an effective guide and travels in the opposite direction.    c) +25 ms: the $m_F=-1$ beam continues to propagate while the $m_F=+1$ beam is turned around and returned toward the origin.   d) +45 ms: the $m_F=+1$ beam now falls freely and experiences mean-field expansion, like the $m_F=-1$ component before it.   Note a slightly different path than $m_F=-1$.   e) +50 ms: continued $m_F=+1$ propagation; note the $m_F=-1$ component has traveled out of the field of view by this point.   Images are 1 mm $\times$ .25 mm; gravity is directed toward the lower right and the trapping laser is directed toward the upper right. }
\end{figure}

After compression-driven spin mixing, we implement magnetic outcoupling, similar in principle to the all-optical $m_F=0$ atom laser reported in 2003\cite{tubingen} but applied in a state-selective fashion whereby applied field gradients nonadiabatically distort the trapping potential and provide a velocity kick for spin-polarized atoms to escape.   Fig.~\ref{twinbeam} illustrates a typical run; here, a magnetic field gradient dominantly directed along the weak axis of the optical trap (toward the upper right) is turned on quickly (over several ms) at a variable delay (5-50 ms) before the optical trap is turned off.   In this case ballistic expansion is limited to a minimum time of 100 $\mu$s (nonzero so as to allow for optical pumping), and is thus effectively \textit{in situ}.

Using slightly different velocity kicks we observe several variants of the dual-beam atom laser.   Most commonly, we observe immediate outcoupling and ballistic flight of the $m_F=-1$ component while the $m_F=+1$ component first propagates in the opposite direction (as expected), reverses its motion, passes through the parent $m_F=0$ condensate, and finally escapes, as depicted in Fig.~\ref{twinbeam}.     With greater velocity kicks, we also observe the more intuitive case of both polarized components escaping into ballistic flight from opposite ends of the cigar-shaped trap.    Rarely, we observe partial exit of an evolved fraction due to insufficient magnetic field tilt compared to the condensate chemical potential $\mu$, which illustrates the fine control of output coupling possible using this scheme.   Deviation from horizontal of the trap itself is also assumed to bias the exit paths of the evolved components.

\begin{figure}[t]
\includegraphics[width=3.375in]{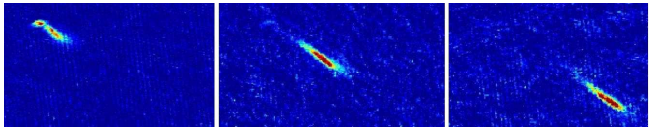}
\caption{\label{downlaser}  Downward outcoupling of the supported $m_F=+1$ condensate (cf. Fig.~\ref{becpalette}c), generated by removing the supportive gradient that had preferentially created a polarized condensate.  This is shown to illustrate the difference in outcoupled beam collimation between the case where mean-field expansion is along the direction of propagation and the case of the spinor dynamics-driven atom laser in Fig. \ref{twinbeam}, where expansion is perpendicular to travel.  Images are 1.2 mm $\times$ .75 mm.}
\end{figure}

The shape of the outcoupled fraction demands close inspection.    To begin with, the transverse atom laser density profile obtained by gravity-directed beams\cite{bloch_laser1999,tubingen} is not to be expected here, as we outcouple over timescales short enough such that the combination of gravity and local gradients do not move the outcoupled fraction appreciably over the time that the trap height $U_0$ tilts significantly with respect to $\mu$.   This short outcoupling time is helpful in making sure that the broad resulting clouds are visible.   We do observe a downward-directed atom laser in the case of the aforementioned `supported' condensate, depicted in Fig.~\ref{downlaser}, outcoupled simply by removing the support in the last few ms before ballistic expansion.    This effect is unrelated to the physics of spin mixing or number-correlated dual-beam atom lasers and is essentially analogous to the T\"{u}bingen experiment\cite{tubingen}.

The immediate broadening of the outcoupled spinor atom laser pulse compared to the downward-directed one is perhaps counterintuitive, but can be explained in terms of the nature of the condensate and the path it takes via the long axis of the trap, rather than the more traditional transverse outcoupling.    The horizontally outcoupled beam experiences preferred mean-field expansion perpendicular to the direction of travel, and thus does not exhibit the tight collimation characteristic of our downward-directed $m_F=+1$ pulse.   We have duplicated this behavior using simple simulations combining center-of-mass motion and standard mean-field expansion theory.

Since not all of the initial reservoir is used up in the creation of the dual-beam atom laser, in principle our compression and outcoupling process could be repeated as long as a population of $m_F=0$ atoms remained.   The current experiment does not produce large condensate numbers, but several groups have produced large-N spinor condensates showing that such quasi-continuous generation of spinor dynamics-driven atom laser pairs is achievable.

The impetus behind this work is the set of proposals that seek to create squeezing and entanglement in oppositely-propagating atom laser beams.    Future work will explore the nature of correlations and entanglement in these beams, the possibility of spin-independent outcoupling as in the 2000 proposals, and also explore the possibilities of improving this process into the quasi-continuous regime.   Also of interest is the possibility of creating undisturbed number states, which would be useful for Heisenberg-limited precision phase measurements\cite{kasevichBEC}.

We acknowledge helpful conversation with Eric Burt, Nan Yu, James Kohel, and Kenneth Libbrecht.  The research described here was carried out at the Jet Propulsion Laboratory, California Institute of Technology, under a contract with the National Aeronautics and Space Administration. 


\end{document}